\documentstyle[12pt]{article}

\newcommand{\beq}{\begin{equation}}
\newcommand{\eeq}{\end{equation}}
\newcommand{\beqa}{\begin{eqnarray}}
\newcommand{\eeqa}{\end{eqnarray}}

\begin{document}

\begin{titlepage}

\begin{flushright}
CERN-TH/97-86\\
UCSD-97-11\\
CALT-68-2114\\
May 1997\\
hep-ph/9705286
\end{flushright}

\vspace{1cm}
\begin{center}
\Large\bf
Kinematic Enhancement of Non-Perturbative Corrections to 
Quarkonium Production 
\end{center}

\vspace{0.7cm}
\begin{center}
{\sc M. Beneke$^a$~~ I.Z. Rothstein$^b$~~ Mark B. Wise$^c$}\\[0.5cm]
{\sl $^a$ Theory Division, CERN, CH-1211 Geneva 23, Switzerland}\\[0.3cm]
{\sl $^b$  Department of Physics University of California at San Diego\\
La Jolla, CA 92122, U.S.A.}\\[0.3cm]
{\sl $^c$ California Institute of Technology Pasadena, CA 91125, 
U.S.A.}\\[0.3cm]
\end{center}

\vspace*{1cm}

\centerline{\bf Abstract}
\vspace*{0.2cm}
\noindent In this letter we address issues involved in quarkonium 
production near the boundaries of phase space. It is shown 
that higher-order non-perturbative contributions are enhanced 
in this kinematic region and lead to a breakdown of the
non-relativistic (NRQCD) expansion. This breakdown is a consequence of  
sensitivity to the kinematics of soft gluon radiation and to the 
difference between partonic and hadronic phase space. We show how
these large corrections can be resummed giving the dominant
contribution to the cross section. The resummation leads to the introduction 
of non-perturbative,  universal distribution functions. We 
discuss the importance of these shape functions for several 
observables, in particular the energy distribution of 
photo-produced $J/\psi$ close to the endpoint.
\end{titlepage}

Non-relativistic effective field theory (NRQCD) \cite{BBL} 
for quarkonium decays and production has shed new light on many 
aspects of the experimental data. In particular, the inclusion 
of color-octet decay and production channels rectified some 
of the main theoretical and phenomenological deficiencies of the 
color-singlet model and could explain the so-called 
`$\psi'$-anomaly' in $p\bar{p}$ collisions at the Tevatron 
\cite{BF95}. Despite this success, the verification of the universality 
of long-distance parameters in NRQCD through the  comparison of 
different production processes turns out to be difficult due 
to significant theoretical uncertainties of various kinds. 
A large uncertainty stems from the treatment
of the kinematics of the hadronization process, which is
especially important near the boundary of phase space where
keeping the leading order terms in the NRQCD expansion is not valid.
Here we concern ourselves exactly with this problem. 
A partial resummation of the NRQCD expansion leads us to introduce 
universal shape functions that parameterize quarkonium formation 
at its kinematic limits. We discuss the effect of these shape 
functions on quarkonium production rates in a variety of production 
processes.  

In the NRQCD approach the inclusive  quarkonium production process factorizes 
into the production of a (heavy) quark-anti-quark pair at small 
relative momentum
\begin{equation}
\mbox{initial state} \to Q\bar{Q}[n] + \tilde{X},
\end{equation}
followed by the `hadronization' of the $Q\bar{Q}$ pair into a 
quarkonium $H$ and light hadrons
\begin{equation}
\label{hadronization}
Q\bar{Q}[n] \to H + X.
\end{equation}
The cross section  may be  written as
\beq 
\label{cross}
d\sigma_H =  \sum_n d\sigma_{Q\bar{Q}[n]+\tilde{X}} \langle  
{\cal O}^H_n \rangle,
\eeq
where $d\sigma_{\bar{Q}Q[n]+\tilde{X}}$ is the short distance cross section to
produce a $Q\bar{Q}$ state labeled by $n$ plus $\tilde{X}$,
and the general form of the NRQCD matrix 
element $\langle  {\cal O}^H_n \rangle$ is
given by
\beq
\label{me}
\langle  {\cal O}^H_n \rangle=\sum_X \,\langle 0|\psi^\dagger \Gamma_n
\chi|H+X \rangle \langle H+X|\chi^\dagger \Gamma^\prime_n \psi
|0 \rangle.
\eeq
$\Gamma_n$ ($\Gamma^\prime_n$) is a matrix in color and spin 
and may contain derivatives and, in principle, gluon and light quark 
fields.

The short-distance cross section is insensitive to 
quarkonium binding by construction  and therefore can depend only on the heavy quark mass 
and kinematic invariants constructed from parton momenta. The quarkonium 
mass never enters explicitly, since any non-perturbative effect is 
parametrized by one of the matrix elements $\langle  {\cal O}^H_n \rangle$. 
As a further consequence of factorization, the momentum taken by the 
light hadronic final state $X$ in (\ref{hadronization}) is neglected 
at lowest order in the velocity expansion \cite{BBL}. This is consistent 
since the energies involved in (\ref{hadronization}) are small, 
of order $m_Q v^2$, in the $Q\bar{Q}$ rest frame, and because the 
process is inclusive over the light hadronic final state. Thus, given 
that the difference between partonic and hadronic kinematics
is higher order in $v^2$, shifting say a partonic
to a hadronic endpoint, does not in general improve
the accuracy of a calculation if other  effects of the same order
in the velocity expansion are left out. 
However, there are circumstances, 
namely near the boundaries of phase space, where effects higher order 
in $v^2$ related to kinematics are enhanced and lead to a breakdown of 
the velocity expansion. In these cases it becomes
necessary to include these effects in a systematic fashion. The breakdown 
of the velocity expansion is closely related to the fact that 
near the boundaries of phase space the details of the hadronization 
process (\ref{hadronization}) are probed at a scale smaller than 
$m_Q v^2$. This happens, for instance, for quarkonium production 
close to threshold, $\hat{s}\to M_H^2$, or in $J/\psi$ photo-production 
close to the point where all the photon's energy is transferred to 
the $J/\psi$ ($z=E_{J/\psi}/E_\gamma\to 1$ in the proton rest frame). 
In these cases, to make 
the NRQCD expansion convergent,
 one must smear sufficiently (in $\hat{s}$ and $z$, 
respectively) so as not to probe the details of hadronization. However,
it is possible to reduce the necessary smearing region by
resumming a class of leading twist operators into universal 
shape functions, thus extending the range of applicability of 
NRQCD. Resummations of related nature have been introduced to 
treat energy spectra in semileptonic or radiative $B$ meson 
\cite{bres} and quarkonium \cite{RW} decays.  

Each kinematic situation must be dealt with
slightly differently, although the general idea is the same.
Consider some general production process $A+B\rightarrow H+\tilde{X}$. 
To determine the matching coefficients  
$d\sigma_{Q\bar{Q}[n]+\tilde{X}}$ at lowest order
we calculate the inclusive rate in full QCD 
to produce a free quark anti-quark pair in a state $n$. The momenta 
of the quark and anti-quark are defined as
\beq
\label{momenta}
p_\mu=\frac{P_\mu}{2}+ (\Lambda  l_1)_\mu \qquad\bar{p}_\mu=
\frac{P_\mu}{2}+ (\Lambda  l_2)_\mu,
\eeq
respectively. $\Lambda$ is the Lorentz
transformation which boosts from the frame where 
$P_\mu$ is given by $(2m_Q,\vec{0})$ to whatever frame one chooses 
for the calculation. In the $P$-rest frame 
$l_i=(\sqrt{\vec{l}_i^2+m_Q^2}-m_Q,\vec{l}_i)$ so that 
$p^2=\bar{p}^2=m_Q^2$. The production cross section is then expanded in 
the small momenta $\vec{l}_i\sim O(m_Q v)$, and powers of momentum 
are identified 
with derivatives acting on heavy quark fields in NRQCD. Thus, a 
factor of relative momentum $l_{1i}-l_{2i}$ corresponds to 
$(\psi^{\dagger} i\rlap{\raise8pt\hbox{$\leftrightarrow$}}{D}_i \chi)$ and a 
factor of center-of-mass (cms) momentum (of the $Q\bar{Q}$ pair in the 
$P$-rest frame) $l_{1i}+l_{2i}$ is identified with a total 
derivative $iD_i (\psi^\dagger \chi)$. Four-fermion operators with 
total derivatives on fermion bilinears are usually ignored since they  
are of higher order in the
non-relativistic expansion than the relative momentum operators. 
They were first written down in \cite{MS} and were later  shown
to be of dynamical importance in radiative quarkonium decay
near the endpoint \cite{RW,MW}. (Note that using the equations of
motion $l_{10}+l_{20}$ can be identified with the total time derivative
$iD_0(\psi^{\dagger} \chi)$.)

It is precisely these cms-derivative operators that cause the 
leading singular contributions in the NRQCD expansion in a general 
situation of constrained phase space. The phase space measure 
for the process, $\mbox{initial state} \to Q(p)+\bar{Q}(\bar{p})+\tilde{X}$, 
is given by
\beq
\label{ps}
d\mbox{PS}=\frac{d^3P}{(2\pi)^3 2P_0}\prod_i 
\frac{d^3k_i}{(2\pi)^3 2k_0}
\,\delta^4\!\left(\sum_l t_l-\sum k_i - (P+\Lambda (l_1+l_2))\right),
\eeq
where the $t_l$ are the incoming momenta and $k_i$ are the momenta of 
final state particles other than the $Q\bar{Q}$ pair. 
The matching strategy then tells us to expand this delta-function, 
as well as the amplitude squared, in powers of the momenta $l_1$ and 
$l_2$. The delta-function depends explicitly on the cms momentum 
$\vec{l}_1+\vec{l}_2$ and implicitly on the relative momentum 
through the zero-components of $l_i$. The expansion of the delta-function 
results in a series in $v^2$, but with increasingly singular 
distributions at the boundary of partonic phase space defined by the 
delta-function with $l_i$ set to zero. Normally, the phase space 
is integrated with a smooth 
function, and the terms coming from the expansion of the delta
function converge rapidly, provided $v^2$ is small. However, if the 
weighting function has support mainly in a region of order $\epsilon m_Q$ 
near the boundary of phase space, the expansion parameter is 
$v^2/\epsilon$. Thus $\epsilon\gg v^2$ is required for convergence. 
On the other hand, resumming the leading terms $(v^2/\epsilon)^n$ to 
all orders allows us to take $\epsilon\sim v^2$. Provided that 
the squared amplitude for the production process is non-singular 
at the boundary of phase space, the resummation of the leading singular 
contributions is easily accomplished. Note that even after resummation 
we can not consider $\epsilon \ll v^2$, because all subleading 
terms $v^{2 m}(v^2/\epsilon)^n$ become important as well in this region.
 Since the details depend on 
the particular kinematic situation, we shall show how this works 
in several processes of topical interest.

Let us begin by considering the case of gluon fragmentation into 
$\psi$ ($J/\psi$ or $\psi'$) through a $Q\bar{Q}$ pair in a 
${}^3\!S_1$ color-octet state, which is now regarded as the 
dominant source of direct $\psi$ production at large transverse 
momentum in hadron collisions \cite{BF95}. The octet 
$Q\bar{Q}$ pair evolves non-perturbatively
into the $\psi$, a process necessarily accompanied by soft gluon 
radiation. At lowest order in $v^2$ the momentum of the soft gluon radiation 
can be neglected and and the fragmentation function is given by
\beq
\label{lwst}
D_{\psi/g}(\hat{z},2m_c)=\frac{\pi \alpha_s(2m_c)}{24 m_c^3}
\delta(1-\hat{z})
\langle{\cal O}_8^\psi({}^3\!S_1)\rangle.
\eeq
The hat reminds us that the variable $\hat{z}=P_+/k_+$ is defined in 
terms of the light-cone $+$-components of parton momenta, that is, 
$P_+$ refers to $P$ in (\ref{momenta}) rather than the $\psi$-momentum. 
$k$ is the momentum of the fragmenting gluon. 
As noted in \cite{bm} this fragmentation function is folded 
with a gluon production cross section, which, for a given value of the 
$\psi$ transverse momentum, is a rapidly varying function
of $\hat{z}$. To be precise   
\begin{equation}
\label{int}
\frac{d\sigma}{dp_t} = \int\limits_{\hat{z}_{min}}^1
\frac{d\hat{z}}{\hat{z}}\,
K(\hat{z},p_t)\,D_{\psi/g}(\hat{z},p_t),
\end{equation}
where $K(\hat{z},p_t)\sim \hat{z}^5$ in the region $p_t\sim 15\,$GeV. 
Consequently, the integral (\ref{int}) 
depends sensitively on the shape of the fragmentation function 
near $\hat{z}=1$ and neglecting the kinematics of soft gluon emission 
is not a good approximation. Intuitively, one expects soft gluon radiation 
to soften the fragmentation function\footnote{This softening occurs 
before the onset of Altarelli-Parisi 
evolution that further softens the fragmentation function at the 
scale $p_t$.} (\ref{lwst}) and therefore to 
reduce $d\sigma/dp_t$ \cite{bm}. However, this intuition relies 
on interpreting $\hat{z}$ as the quarkonium $+$-momentum fraction,  
different from the definition of $\hat{z}$ given above.

Now the existence of cms momentum $l=l_1+l_2$ in the matching relations 
can be interpreted in the quarkonium rest frame as recoil of the 
$Q\bar{Q}$ pair against the emitted soft gluons. Indeed, since the 
matrix element for $g^*\to Q\bar{Q}$ is evidently non-singular at 
$\hat{z}=1$, the leading singular contributions to the 
coefficient functions of higher-dimension operators in the 
NRQCD expansion simply follow from (\ref{ps}). After integrating 
over phase space except for the $+$-direction, this leads to the
shift
\beq
\delta(1-\hat{z})\rightarrow \delta(1-\hat{z}-\Lambda_{+\mu}l^\mu/P_+).
\eeq
In terms of  operators in the NRQCD expansion, 
expanding the modified 
delta-function in $\Lambda_{+\mu}l^\mu /P_+=l_+/(2 m_Q)$, 
we obtain the series 
\begin{eqnarray}
\label{sers}
D_{\psi/g}(\hat{z},2m_c)&=&\frac{\pi \alpha_s(2m_c)}{24 m_c^3}
\sum_m \frac{1}{m!}\delta^{(m)}(1-\hat{z})
\nonumber\\
&&\,\sum_X\langle 0 | \psi^\dagger \sigma_i T^A \chi |\psi+X \rangle 
\langle \psi+X |(i n\cdot \hat D)^m \left(\chi^\dagger 
\sigma_i T^A \psi\right) | 0\rangle,
\end{eqnarray}
where $n$ is the light like vector $(1,0,0,-1)$, $\hat D=D/(2m_c)$
and $\delta^{(m)}(1-\hat{z})$ denotes the m'th derivative of
the delta-function. The first term 
in the sum reproduces (\ref{lwst}). Introducing the  shape function
\beq
\label{fragdist}
F_{\psi}[{}^3\!S_1^{(8)}](y_+)=\sum_X \langle 0 | \psi^\dagger
\sigma_iT^A \chi 
|\psi+X\rangle \langle \psi+X |\delta\!\left(y_+-in\cdot \hat D \right)
\left(\chi^\dagger \sigma_i T^A \psi \right) |0 \rangle,
\eeq
we rewrite (\ref{sers}) as 
\beq 
\label{sh}
D_{\psi/g}(\hat{z},2m_c)=\frac{\pi \alpha_s(2m_c)}{24 m_c^3}
\int dy_+ \delta(1-\hat{z}-y_+)\,F_{\psi}[{}^3\!S_1^{(8)}](y_+).
\eeq
Note that $F_{\psi}[{}^3\!S_1^{(8)}](y_+)$ is a non-perturbative 
distribution function that can not be calculated except within models. 
But the factorization as expressed by (\ref{fragdist}) and 
(\ref{sh}) implies that such shape functions are process-independent. 
Moreover, $F_{\psi}[{}^3\!S_1^{(8)}](y_+)$ has support mainly 
for $|y_+| < v^2$. We assume 
$m_Q v\gg m_Q v^2\sim \Lambda_{QCD}$ for NRQCD power counting, in which 
case a cms-derivative acting on a quark bilinear scales as 
$m_Q v^2$, the momentum of dynamical gluons in NRQCD.\footnote{In the 
construction of \cite{GR} this can be seen from the fact that the 
matrix element with $n$ cms derivatives vanishes
unless one inserts an n'th multipole operator resulting in a 
an additional $1/c^n$ factor above those coming from naive dimensional
analysis.} It is 
suggestive to interpret $F_{\psi}[{}^3\!S_1^{(8)}](y_+)$ as the probability 
for soft gluons to carry off 
a light-cone momentum fraction $y_+$ in the hadronization of the 
color-octet $Q\bar{Q}$ pair. Although useful, this interpretation 
is not quite exact, because, as will be seen in more detail below, 
such distribution functions are non-trivial even for 
color-singlet $Q\bar{Q}$ states, when no soft gluon emission is 
required. In such cases, the shape functions account for the 
difference between quark and quarkonium masses and the corresponding 
difference in the definition of kinematic variables. 

The $\psi$ production rate at fixed $p_t$ follows from integrating the
fragmentation function with a function that roughly varies as 
$\hat{z}^4$, see (\ref{int}). Eq.~(\ref{sers}) implies that the 
NRQCD expansion parameter is then $4 v^2\sim 1$ rather than 
$v^2$. Inclusion of the effects
due to the shape function $F_{\psi}[{}^3\!S_1^{(8)}](y_+)$ 
can therefore alter the $\psi$ production cross section of order unity, 
an effect that should be kept in mind when 
$\langle {\cal O}_8^\psi({}^3\!S_1)\rangle$ extracted from 
Tevatron data \cite{CL} is compared with the extraction from another 
production process. 
Unfortunately, without a non-perturbative
calculation of the shape function, no detailed prediction is possible. 
We note that the prediction that the $\psi$ is transversely
polarized at large $p_t$ \cite{CW} in hadron collisions remains 
unaffected by the 
resummation procedure described here. (For perturbative corrections to
this prediction see \cite{BR1}.) 

Our second example concerns the total hadro-production cross section 
of a quarkonium $H$ at fixed target energies. In this case we write the
cross section as
\beq
\label{fact}
\sigma_H = \sum_{i,j}\int\limits_0^1 d x_1 d x_2\,
f_{i/A}(x_1) f_{j/B}(x_2)\,\hat{\sigma}(ij\rightarrow H+\tilde{X})\,,
\end{equation}
The sum extends over all partons in the colliding 
hadrons and $f_{i/A}$, etc. denote the corresponding parton distribution 
functions. The parton cross section 
$\hat{\sigma}$ is a distribution that is weighted with the 
parton densities. Let us consider the leading parton process 
$i+j\to Q(p)+\bar{Q}(\bar{p})$. Since its matrix element is again 
non-singular at the boundaries of phase space, it is sufficient 
to retain $l_i$ in (\ref{ps}) and to set $l_i=0$ in the matrix element, 
in order to obtain the most singular coefficient functions to all 
orders in $v^2$. The parton cross section may then be 
written as 
\beqa
\label{had21}
\hat \sigma(ij\rightarrow H) &=& \frac{\pi^3 \alpha_s^2(2m_Q)}{(2m_Q)^3} 
\sum_n\,C^{ij}[n]\,\nonumber\\
&&\hspace*{-2.5cm}
\sum_X\langle 0 |\psi^\dagger \Gamma_n
\chi|H+X\rangle\langle H+X |\delta (x_1 x_2 s-4 m_Q^2-
2 (k_i+k_j)\cdot\Lambda l)\,\chi^\dagger 
\Gamma^\prime_n \psi |0 \rangle ,
\eeqa
where $k_{i,j}$ are the four-momenta of the incoming partons, 
$s$ the hadron center-of-mass energy and $l=i D$. The sum over 
$n$ extends over all possible operators and is truncated according
to the non-relativistic power counting rules. For the two most 
important production channels in $S$-wave quarkonium 
production $C^{gg}[{}^1\!S_0^{(8)}]=5/12$, 
$C^{gg}[{}^3\!P_0^{(8)}]=35/(12 m_Q^2)$ \cite{BR2,others}. Using 
$2 (k_i+k_j)\cdot\Lambda l\approx 2 P\cdot\Lambda l=4 m_Q l^0$
and expanding the delta-function gives the series of higher-dimension 
operators
\beqa
\label{had31}
\hat\sigma(ij \rightarrow H)&=& \frac{\pi^3 \alpha_s^2(2m_Q)}{(2m_Q)^3s} 
\sum_n\,C^{ij}[n]\,\sum_m\,\frac{1}{m!}\delta^{(m)}
(x_1x_2-\frac{4m_Q^2}{s}) \nonumber \\
&&\hspace*{-1cm}
\sum_X\langle 0|\psi^\dagger \Gamma_n\chi|H+X\rangle\langle H+X|
\left(\frac{8 m_Q^2 i\hat{D}_0}{s}\right)^{\!m} \!\left(\chi^\dagger 
\Gamma^\prime_n \psi\right)|0 \rangle, 
\eeqa
which accounts for all most singular contributions at $x_1 x_2=4 m_Q^2/s$. 
As previously, $\hat{D}=D/(2 m_Q)$. Note that as opposed to (\ref{sers}), 
the higher-dimension operators  involve time rather than +-derivatives. 
Accordingly we define a shape function in energy fraction $y_E$
\beq
\label{haddist}
F_{H}[n](y_E)=\sum_X \langle 0 | \psi^\dagger
\Gamma_n\chi 
|H+X\rangle \langle H+X |\delta\!\left(y_E-i\hat D_0 \right)
\left(\chi^\dagger \Gamma_n^\prime\psi \right) |0 \rangle.
\eeq
We emphasize that $F_{H}[n](y_E)$ and $F_{H}[n](y_+)$ in 
(\ref{fragdist}) are different non-perturbative distributions. 
Up to the caveat mentioned above, $F_{H}[n](y_E)$ can be 
interpreted, in the quarkonium rest frame, as the 
distribution of energy fraction taken by soft gluons during the
transition from the short-distance heavy quark pair in a state $n$ 
into the quarkonium $H$.

Consider now the case where the $Q\bar{Q}$ pair is in the same 
color-singlet state as the quarkonium $H$. 
Up to corrections suppressed in $v^2$, we can use the 
vacuum saturation approximation and drop the sum over $X$ in 
(\ref{haddist}). The derivative $iD_0$ produces a factor of
the binding energy and we obtain
\beq
\label{vacsat}
F_{H}[n](y_E)=\delta\!\left(y_E-\frac{M_H-2 m_Q}{2 m_Q}\right)\,
\langle 0 | \psi^\dagger
\Gamma_n\chi |H \rangle \langle H |
\chi^\dagger \Gamma_n^\prime\psi |0 \rangle,
\eeq
where $M_H$ is the quarkonium mass. Consequently, the $y_E$-integral 
that enters the production cross section (see (\ref{resum2}) below) 
can be done:
\begin{eqnarray}
\int \!dy_E\,F_{H}[n](y_E)\,\delta(x_1 x_2 s-4 m_Q^2-8 m_Q^2 y_E) 
&= &\delta\!\left(x_1 x_2 s-4 m_Q^2-4 m_Q (M_H-2 m_Q)\right) \nonumber\\
&&\hspace*{-1cm}
\cdot \langle 0 | \psi^\dagger
\Gamma_n\chi |H \rangle \langle H |
\chi^\dagger \Gamma_n^\prime\psi |0 \rangle.
\end{eqnarray}
Since $M_H^2=4 m_Q^2+4 m_Q (M_H-2 m_Q)+O(v^4)$, we see that 
the sole effect of the distribution function in the vacuum 
saturation approximation is to shift the 
unphysical partonic boundary of phase space to the hadronic one. 
The inclusion of sub-leading terms would complete the
transmutation 
\beq
\label{resum1}
\delta(x_1 x_2 s-4m_Q^2)\rightarrow \delta(x_1 x_2 s-M_H^2).
\eeq

As we previously mentioned, performing this resummation does
not in general improve the accuracy of the calculation, given that
other contributions of the same order in $v^2$ have been left out. 
However, if the corrections from the expansion
of the delta-function are enhanced because the dominant contribution 
to the cross section comes from a region close to the boundary of phase 
space, then we have indeed improved the
situation. Introducing the variable $\tau = x_1x_2$ and the 
parton luminosities $L_{ij}(\tau)=\int_\tau^1 dx/x\,f_{i/A}(x) 
f_{j/B}(\tau/x)$, the hadro-production cross section can be 
expressed as 
\beq
\label{resum2}
\sigma_H = \frac{\pi^3\alpha_s^2(2m_Q)}{(2m_Q)^3}
 \sum_{i,j}\,\int_0^1 \!d\tau\,L_{ij}(\tau)\,\sum_n \,C^{ij}[n]
\int \!dy_E \,F_{H}[n](y_E)\,\delta (\tau s-4m_Q^2-8m_Q^2 y_E).
\eeq
At high energies $s\gg 4 m_Q^2$, the gluon-gluon fusion channel 
dominates and we see that the partonic cross section is weighted 
by a gluon luminosity, that, because of the small-$x$ behaviour of 
the gluon distribution, increases rapidly as $\tau$ decreases. 
Thus, a systematic shift in $\tau$ due to an asymmetric 
shape function $F_{H}[n](y_E)$ (such as the example (\ref{vacsat}) 
above) can affect the magnitude of the total cross section in 
an important way. Indeed, one finds in the leading order 
in $v^2$ predictions that the difference between 
quark and quarkonium masses in the phase space delta-function 
introduces a normalization uncertainty of up to a factor two 
for $\psi'$ and $\chi$ production. As noted in \cite{BR2}, 
the effect 
should be even more pronounced in color-octet mechanisms as, due 
to soft gluon radiation, the final state invariant mass is always 
larger than $M_H^2$. 

As our final, and perhaps most interesting, example we examine 
photo-production of $J/\psi$, in particular the $z$-distribution, 
where $z=p\cdot P_{J/\psi}/p\cdot k_\gamma$ and $p$ is the proton 
momentum.\footnote{Because we will be concerned with the 
large-$z$ region, we do not consider resolved photon interactions 
in the following.} ($z=E_{J/\psi}/E_\gamma$ in the proton rest frame.) To 
exclude higher-twist contributions from diffractive $J/\psi$ 
production, which can not be accommodated in the leading-twist 
NRQCD factorization formalism, it is preferable to consider 
inelastic $J/\psi$ production and to impose $z<0.9$ and $P_t>1\,$GeV. 
Inelastic $J/\psi$ photo-production is often perceived as a problem 
for the NRQCD theory of onium production, because the color-octet 
contributions to this process rise rapidly with $z$ \cite{photo}, 
which is not supported by the data. On the other hand  
octet production is accompanied by gluon radiation, which carries 
away an energy fraction of order $v^2$ and makes it highly unlikely 
to reach a value of $z$ close to one \cite{BR2}. Technically, 
the necessary smearing of the energy distribution is again achieved 
through a shape function, as we shall see, although the situation 
is slightly more complicated for the $2\to 3$ parton processes than 
for the $2\to 2$ parton processes considered before.

Before addressing this issue, let us briefly comment on the 
$2\to 2$ parton processes in photo-production, which are kinematically 
restricted to $\hat{z}=p\cdot P/p\cdot k_\gamma=1$ and $P_t=0$. 
We emphasize again that the calculation in NRQCD refers to 
partonic variables like $\hat{z}$ with $P$ defined 
in (\ref{momenta}) rather than $z$. If we are interested in the 
total cross section then the analysis goes through almost exactly 
as in the case of hadro-production. In fact the same shape function 
(\ref{haddist}) enters. However, the total cross section is of 
marginal interest experimentally. More interesting are the 
differential distributions. For the gluon fusion process
\beqa
\label{photo}
\frac{d\sigma_H}{d\hat{z}d^2P_t} &=& \frac{\pi^3\alpha_s(2m_Q)\alpha_{em}}
{\hat{z} (2m_Q)^3}\!\int\limits_0^1\!dx\,f_{g/P}(x)
\int \!dy_E dy_+ d^2\rho_t \sum_n\,C^g[n] F_{H}[n](y_E,y_+,\vec{\rho}_t) 
\nonumber\\
&&\,\delta(sx-4m_Q^2-8m_Q^2 y_E)\,\delta(1-\hat{z}-y_+)\,
\delta^{(2)}(\vec P_t-\vec\rho_t),
\eeqa
where 
\beqa
F_{H}[n](y_E,y_+,\vec{\rho}_t)&=&\sum_X\langle 0|\psi^\dagger \Gamma_n
\chi|H+X \rangle\nonumber\\
&&\hspace*{-2cm} \langle H+X|\delta(y_+-i n\cdot\hat{D})
\delta(y_E-i \hat{D}^0)\delta(\vec\rho_T - i\vec D_T)
\left(\chi^\dagger \Gamma^\prime_n \psi\right)|0 \rangle.
\eeqa
Such a multi-dimensional distribution is much too complicated to be 
useful in practice, and we do not pursue it further here. Qualitatively, 
this distribution causes smearing in $\hat{z}$ and transverse momentum 
over a region (for $Q=c$)
\beq
\label{reg}
\delta \hat{z} \approx v^2 \approx 0.25 - 0.3, \qquad 
\delta P_t \approx  m_c v^2 \approx 0.5\,\mbox{GeV}.
\eeq
Unless the endpoint region is averaged over an interval of this 
size, the NRQCD expansion fails to be convergent, even if the above 
distribution function is taken into account. Smearing over a range 
much larger than (\ref{reg}), makes the corrections considered here 
irrelevant and one can return to the ordinary velocity expansion. 
Note that because of the sizeable smearing in $\hat{z}$, 
part of the $2\to 2$ parton contributions that are normally considered 
to be localized at $z=1$ actually leaks into the inelastic region 
$z<0.9$. This leakage can be eliminated by a transverse 
momentum cut $P_t\gg m_c v^2$. Note also that the smearing in the transverse 
direction is of the same order of magnitude as the smearing 
caused by intrinsic transverse momentum of the gluon in the proton. 

Let us return now to inelastic photo-production of quarkonium and 
consider $\gamma+g\to Q(p)+\bar{Q}(\bar{p})+g(k)$. It is 
straightforward to obtain
\beqa
\label{photocross}
\frac{d\sigma}{d\hat{z}} &=& \int\limits_{P^2_{t,min}}\!dP_t^2
\int\limits_0^1\!dx\,S(x,\hat{z},P_t^2)\,
\sum_X\langle 0|\psi^\dagger \Gamma_n
\chi|H+X \rangle
\nonumber\\
&&\,\langle H+X|\delta\!\left(s(1-\hat{z}) x-\frac{M^2 (1-\hat{z})+P_t^2}
{\hat{z}}-2 k\cdot\Lambda iD\right)
\left(\chi^\dagger \Gamma^\prime_n \psi\right)|0 \rangle,
\\
&&\hspace*{-1.2cm}
S(x,\hat{z},P_t^2) \,=\, \frac{1}
{16\pi \hat{z} xs}\,f_{g/p}(x)\,|{\cal M}_n|^2,
\nonumber
\eeqa
where $s$ is the cms energy, $M=2 m_Q$, and ${\cal M}_n$ is the matrix element 
to produce the $Q\bar{Q}$ pair in a state $n$. If we neglect the 
covariant derivative $D$ in the delta-function, the standard kinematic 
relations for the inelastic processes considered in \cite{photo} 
are recovered. To proceed, we have to extract the dependence of 
$2k\cdot \Lambda D$ on $\hat{z}$ and $P_t$. To this end we 
note that 
\beqa
\sum_X\langle 0|\psi^\dagger \Gamma_n
\chi|H+X \rangle
\langle H+X|(i D_{\mu_1})\ldots (i D_{\mu_m})
\left(\chi^\dagger \Gamma^\prime_n \psi\right)|0 \rangle 
&=&
\nonumber\\
&&\hspace*{-10cm} 
A_m[n]\,w_{\mu_1}\ldots w_{\mu_m} + g_{\mu_i\mu_j}\mbox{-terms}
\eeqa
in a frame where the $Q\bar{Q}$-pair moves with four-velocity $w$. Since 
$k$ is a light-like four-vector, terms involving
the metric tensor $g_{\mu_i\mu_j}$ do not 
contribute and we obtain, in the rest frame of $P$ (see (\ref{momenta})), 
\beq
\sum_X\langle 0|\psi^\dagger \Gamma_n
\chi|H+X \rangle
\langle H+X|(2 k\cdot \Lambda i D)^m
\left(\chi^\dagger \Gamma^\prime_n \psi\right)|0 \rangle 
= A_m[n] \,(2 k^0_{rest})^m.
\eeq
Using
\beq
2 k^0_{rest} = \frac{2 P\cdot k}{M} = \frac{M^2 (1-\hat{z})^2+P_t^2}{M 
\hat{z} (1-\hat{z})},
\eeq
we can write (\ref{photocross}) as 
\beqa
\frac{d\sigma}{d\hat{z}} &=& \int\limits_{P^2_{t,min}}\!dP_t^2
\int\limits_0^1\!dx\int\!dy_+\,S(x,\hat{z},P_t^2)\,F_H[n](y_+)
\nonumber\\
&&\hspace*{1cm}
\delta\!\left(s(1-\hat{z}) x-\frac{M^2 (1-\hat{z})+P_t^2}
{\hat{z}}-\frac{M^2 (1-\hat{z})^2+P_t^2}{\hat{z} (1-\hat{z})}
y_+\right),
\eeqa
where the shape function $F_H[n](y_+)$ is defined as in 
(\ref{fragdist}), generalized to an arbitrary $Q\bar{Q}$ state $n$. 
Now let us examine the delta-function. For any non-zero 
$P_t$ we see that the expansion parameter close to $\hat{z}=1$ is 
$y_+/(1-\hat{z})\sim v^2/(1-\hat{z})$. Thus, the 
NRQCD expansion breaks down for $1-\hat{z} \sim O(v^2)$, because higher-order 
terms in $v^2$ grow more and more rapidly 
as $\hat{z}\to 1$.\footnote{The integral over $\hat{z}$ still exists, 
because for 
any fixed $P_t$, $\hat{z}$ can never reach one. However, 
$1-\hat{z}_{max}=O(M^2/s,P_t^2/s)$ is very small in a high energy 
collision. We also mention that in the case considered here, the 
expansion of the partonic amplitude in $l_1$ and $l_2$ can also 
produce terms more singular in $1/(1-\hat{z})$ in higher orders 
in $l_i$.} Consequently, the NRQCD factorization approach makes no 
prediction in the endpoint region and the discrepancy between 
leading order predictions \cite{photo} and data in this region does 
not allow us to draw any conclusion on the relevance of 
color-octet contributions to photo-production. 
If one averages the $\hat{z}$-distribution over a sufficiently 
large region containing the endpoint, the octet mechanisms contribute 
significantly to this average. However, the characteristic shape 
information is then lost and one has to deal with the more difficult 
and uncertain question of whether the absolute magnitude of the 
cross section requires the presence of octet contributions and 
whether their magnitude is consistent with other production processes.
Almost identical remarks apply to the endpoint region in the 
$J/\psi$ energy distribution in the process $B\to J/\psi+X$ or 
$e^+ e^-\to J/\psi+X$.

In conclusion, we have shown how the NRQCD expansion of quarkonium 
production processes breaks down in regions of phase space, where the 
prediction is sensitive either to the difference between parton and 
hadron kinematics or to momentum carried away by light hadrons 
in the hadronization process (\ref{hadronization}). The amount 
of smearing which is necessitated by this breakdown can be reduced 
by introducing universal shape functions in NRQCD as illustrated 
in this letter. Being universal, these functions could in 
principle be extracted from one production process and used 
for another. However, the existence of such functions for 
every $Q\bar{Q}$ state $n$ at short distances, together 
with functions in different variables (such as $y_+$ and $y_E$) 
makes this difficult in practice. 
At this stage it seems worthwhile to use the operator expressions 
for the shape functions to invent phenomenological models for these 
functions, which 
are consistent with the properties of NRQCD.\\

\noindent {\bf Acknowledgements.} 
M.B. wishes to thank the CalTech theory group for 
its hospitality while part of this work was done. M.B.W. was supported in 
part by the U.S. Department of Energy under grant number DE-FG03-92-ER40701.
I.Z.R. was supported in part by the U.S. Department of Energy under grant
number DE-FG03-90-ER40546.

\end{document}